\documentclass[prd,showpacs]{revtex4}
\begin{document}

\newcommand{\re}{\mathop{\mathrm{Re}}}

\newcommand{\be}{\begin{equation}}
\newcommand{\ee}{\end{equation}}
\newcommand{\bea}{\begin{eqnarray}}
\newcommand{\eea}{\end{eqnarray}}

\title{The canonical superenergy tensors and stability of the solutions to the Einstein equations}

\author{Janusz Garecki}
\email{garecki@wmf.univ.szczecin.pl}
\affiliation{\it Institute of Mathematics University of Szczecin
and Cosmology Group University of Szczecin,
 Wielkopolska 15, 70-451 Szczecin, Poland}
\date{\today}
\input epsf
\begin{abstract}
Here we present a new method to study stability of the solutions
to the Einstein equations. This method uses the canonical
superenergy tensors which have been introduced in past in our
papers.
\end{abstract}
\maketitle
\section{Introduction}
In the paper we propose a new approach to study stability of a solution to the Einstein equations.
This approach uses the canonical superenergy tensors which were
introduced into general relativity in our papers \cite {Gar1}.
Namely, we assert that when the total superenergy density, matter
and gravitation, $\epsilon_s$, is non-negative, i.e., when $\epsilon_s\geq
0$, then the solution is stable. Contrary, when $\epsilon_s$ is
negative-definite, i.e., when $\epsilon_s<0$, then the solution is
unstable.

The paper is organized as follows. In Section II we remind problems
with local energy-momentum in general relativity and our proposition to
avoid them -- the canonical superenergy tensors.

In Section III we give examples of an intriguing correlation between
stability of the very known solutions to the Einstein equations
and sign of the total canonical superenergy density, $\epsilon_s$,
for them.
 We claim there that these exciting correlations are consequences of the Proposition,
which we have formulated and proved in this Section.

Finally, the short Section IV, contains our conclusion.

In Appendix we present some results of the last our calculations.

In the paper we use the same signature and notation as used in the
last editions of the famous book by Landau and Lifshitz.
\section{The canonical superenergy tensors}
In the framework of general relativity ({\bf GR}), as a
consequence of the Einstein Equivalence Principle ({\bf EEP}), the
gravitational field {\it has non-tensorial strengths} $\Gamma^i_{kl}
 = \{^i_{kl}\}$ and {\it admits no energy-momentum tensor}. One
 can only attribute to this field {\it gravitational
 energy-momentum pseudotensors}. The leading object of such a kind
 is the {\it canonical gravitational energy-momentum pseodotensor}
 $_E t_i^{~k}$ proposed already in past by Einstein. This
 pseudotensor is a part of the {\it canonical energy-momentum
 complex} $_E K_i^{~k}$ in {\bf GR}.

The canonical complex $_E K_i^{~k}$ can be easily obtained by
rewiriting Einstein equations to the superpotential form
\begin{equation}
_E K_i^{~k} := \sqrt{\vert g\vert}\bigl( T_i^{~k} + _E
t_i^{~k}\bigr) = _F U_i^{~[kl]}{}_{,l}
\end{equation}
where $T^{ik} = T^{ki}$ is the symmetric energy-momentum tensor for matter, $g = det[g_{ik}]$,
 and

\begin{eqnarray}
_E t_i^{~k}& =& {c^4\over 16\pi G} \bigl\{\delta_i^k
g^{ms}\bigl(\Gamma^l_{mr}\Gamma^r_{sl} -
\Gamma^r_{ms}\Gamma^l_{rl}\bigr)\nonumber\cr
&+& g^{ms}_{~~,i}\bigl[\Gamma^k_{ms} - {1\over 2}
\bigl(\Gamma^k_{tp}g^{tp} -
\Gamma^l_{tl}g^{kt}\bigr)g_{ms}\nonumber\cr
&-& {1\over 2}\bigl(\delta^k_s \Gamma^l_{ml} +
\delta^k_m \Gamma^l_{sl}\bigr)\bigr]\bigr\};
\end{eqnarray}
\begin{equation}
_F {U_i^{~[kl]}} = {c^4\over 16\pi G}g_{ia}({\sqrt{\vert
g\vert}})^{(-1)}\bigl[\bigl(-g\bigr)\bigl(g^{ka} g^{lb} - g^{la}
g^{kb}\bigr)\bigr]_{,b}.
\end{equation}
$_E t_i^{~k}$ are components of the canonical energy-momentum
pseudotensor for gravitational field $\Gamma ^i_{kl} =
\bigl\{^i_{kl}\bigr\}$, and $_F {U_i^{~[kl]}}$ are von Freud
superpotentials.
\begin{equation}
_E K_i^{~k} = \sqrt{\vert g\vert}\bigl(T_i^{~k} + _E
t_i^{~k}\bigr)
\end{equation}
are components of the {\it Einstein canonical energy-momentu complex,
for matter and gravity}, in {\bf GR}.

In consequence of (1) the complex $_E K_i^{~k}$satisfies local
conservation laws
\begin{equation}
{_E K_i^{~k}}_{,k}\equiv 0.
\end{equation}
In very special cases one can obtain from these local conservation
laws the reasonable integral conservation laws.

Despite that one can easily introduce in {\bf GR} {\it the
canonical (and others) superenergy tensor} for gravitational
field. This was done in past in a series of our articles (See,
e.g.,\cite{Gar1} and references therein).
It appeared that the idea of the superenergy tensors is universal:
to any physical field having an energy-momentum tensor or
pseudotensor one can attribute the coresponding superenergy
tensor.

So, let us give a short reminder of the general, constructive
definition of the superenergy tensor $S_a^{~b}$ applicable to
gravitational field and to any matter field. The definition uses
{\it locally Minkowskian structure} of the spacetime in {\bf GR}
and, therefore, it fails in a spacetime with torsion, e.g., in Riemann-Cartan
spacetime.
In the normal Riemann coordinates {\bf NRC(P)} we define
(pointwiese)
\begin{equation}
S_{(a)}^{~~~(b)}(P) = S_a^{~b} :=(-) \displaystyle\lim_{\Omega\to
P}{\int\limits_{\Omega}\biggl[T_{(a)}^{~~~(b)}(y) - T_{(a)}^{
~~~(b)} (P)\biggr]d\Omega\over 1/2\int\limits_{\Omega}\sigma(P;y)
d\Omega},
\end{equation}
where
\begin{eqnarray}
T_{(a)}^{~~~(b)}(y) &:=& T_i^{~k}(y)e^i_{~(a)}(y)
e_k^{~(b)}(y),\nonumber\cr
T_{(a)}^{~~~(b)}(P)&:=& T_i^{~k}(P) e^i_{~(a)}(P)e_k^{~(b)}(P) =
T_a^{~b}(P)
\end{eqnarray}
are {\it physical or tetrad components} of the pseudotensor or
tensor field which describes an energy-momentum distribution, and $\bigl\{y^i\bigr\}$
are normal coordinates. $e^i_{~(a)}(y), e_k^{~(b)} (y)$ mean an
orthonormal tetrad $e^i_{~(a)}(P) = \delta_a^i$ and its dual $e_k^{~(a)}(P) = \delta_k^a $
paralelly propagated along geodesics through $P$ ($P$ is the origin
of the {\bf NRC(P)}).
We have
\begin{equation}
e^i_{~(a)}(y) e_i^{~(b)}(y) = \delta_a^b.
\end{equation}
For a sufficiently small 4-dimensional domain $\Omega$ which
surrounds {\bf P} we require
\begin{equation}
\int\limits_{\Omega}{y^i d\Omega} = 0, ~~\int\limits_{\Omega}{y^i
y^k d\Omega} = \delta^{ik} M,
\end{equation}
where
\begin{equation}
M = \int\limits_{\Omega}{(y^0)^2 d\Omega} =
\int\limits_{\Omega}{(y^1)^2 d\Omega} =
\int\limits_{\Omega}{(y^2)^2
d\Omega}=\int\limits_{\Omega}{(y^3)^2 d\Omega},
\end{equation}
is a common value of the moments of inertia of the domain $\Omega$
with respect to the subspaces $y^i = 0,~~(i= 0,1,2,3)$.
We can take as $\Omega$, e.g., a  sufficiently small analytic ball centered
at $P$:
\begin{equation}
(y^0)^2 + (y^1)^2 + (y^2)^2 + (y^3)^2 \leq R^2,
\end{equation}
which for an auxiliary positive-definite metric
\begin{equation}
h^{ik} := 2 v^i v^k - g^{ik},
\end{equation}
can be written in the form
\begin{equation}
h_{ik}y^i y^k \leq R^2.
\end{equation}
A fiducial observer {\bf O} is at rest at the beginning {\bf P}
of the used Riemann normal coordinates {\bf NRC(P)} and its four-
velocity is $v^i =\ast~ \delta^i_o.$ $=\ast$ means that an
equations is valid only in special coordinates.
$\sigma(P;y)$ denotes the two-point {\it world function}
introduced in past by J.L. Synge \cite{Synge}
\begin{equation}
\sigma(P;y) =\ast {1\over 2}\bigl(y^{o^2} - y^{1^2} - y^{2^2}
-y^{3^2}\bigr).
\end{equation}
The world function $\sigma(P;y)$ can be defined covariantly by the
{\it eikonal-like equation} \cite{Synge}
\begin{equation}
g^{ik} \sigma_{,i} \sigma_{,k} = 2\sigma,
~~\sigma_{,i} := \partial_i\sigma,
\end{equation}
together with
\begin{equation}
\sigma(P;P) = 0, ~~\partial_i\sigma(P;P) = 0.
\end{equation}
The ball $\Omega$ can also be given by the inequality
\begin{equation}
h^{ik}\sigma_{,i} \sigma_{,k} \leq R^2.
\end{equation}
Tetrad components and normal components are equal at {\bf P}, so,
we will write the components of any quantity attached to {\bf P}
without tetrad brackets, e.g., we will write $S_a^{~b}(P)$
instead of $S_{(a)}^{~~~(b)}(P)$ and so on.

If $T_i^{~k}(y)$ are the components of an energy-momentum tensor
of matter, then we get from (5)
\begin{equation}
_m S_a^{~b}(P;v^l) = \bigl(2{\hat v}^l {\hat v}^m - {\hat g}^{lm}\bigr) \nabla_l \nabla_m {}
{\hat T}_a^{~b} = {\hat h}^{lm}\nabla_l \nabla_m {}{\hat T}_a^{~b}.
\end{equation}
Hat over a quantity denotes its value at {\bf P}, and $\nabla$
means covariant derivative.
Tensor $_m S_a^{~b}(P;v^l)$ is {\it the canonical superenergy tensor for matter}.

For the gravitational field, substitution of the canonical
Einstein energy-momentum pseudotensor as $T_i^{~k}$ in (5) gives
\begin{equation}
_g S_a^{~b}(P;v^l) = {\hat h}^{lm} {\hat W}_a^{~b}{}_{lm},
\end{equation}
where
\begin{eqnarray}
{W_a^{~b}}{}_{lm}&=& {2\alpha\over 9}\bigl[B^b_{~alm} +
P^b_{~alm}\nonumber\cr
&-& {1\over 2}\delta^b_a R^{ijk}_{~~~m}\bigl(R_{ijkl} +
R_{ikjl}\bigr) + 2\delta_a^b{\beta}^2 E_{(l\vert g}{}E^g_{~\vert
m)}\nonumber\cr
&-& 3 {\beta}^2 E_{a(l\vert}{}E^b_{~\vert m)} + 2\beta
R^b_{~(a\vert g\vert l)}{}E^g_{~m}\bigr].
\end{eqnarray}
Here $\alpha = {c^4\over 16\pi G} = {1\over 2\beta}$, and
\begin{equation}
E_i^{~k} := T_i^{~k} - {1\over 2}\delta_i^k T
\end{equation}
is the modified energy-momentum tensor of matter \footnote{In
terms of $E_i^{~k}$ Einstein equations read $R_i^{~k} = \beta
E_i^{~k}$.}.
On the other hand
\begin{equation}
B^b_{~alm} := 2R^{bik}_{~~~(l\vert}{}R_{aik\vert m)}-{1\over
2}\delta_a^b{} R^{ijk}_{~~~l}{}R_{ijkm}
\end{equation}
are the components of the {\it Bel-Robinson tensor} ({\bf BRT}),
while
\begin{equation}
P^b_{~alm}:= 2R^{bik}_{~~~(l\vert}{}R_{aki\vert m)}-{1\over
2} \delta_a^b{}R^{jik}_{~~~l}{}R_{jkim}
\end{equation}
is the Bel-Robinson tensor with  ``transposed'' indices $(ik)$.
Tensor $_g S_a^{~b}(P;v^l)$ is the {\it canonical superenergy
tensor} for gravitational field $\bigl\{^i_{kl}\bigr\}$.
In vacuum $_g S_a^{~b}(P;v^l)$ takes the simpler form
\begin{equation}
_g S_a^{~b}(P;v^l) = {8\alpha\over 9} {\hat h}^{lm}\bigl({\hat
C}^{bik}_{~~~(l\vert}{}{\hat C}_{aik\vert m)} -{1\over
2}\delta_a^b {\hat C}^{i(kp)}_{~~~~~(l\vert}{}{\hat C}_{ikp\vert
m)}\bigr).
\end{equation}
Here $C^a_{~blm}$ denote components of the {\it Weyl tensor}.

Some remarks are in order:
\begin{enumerate}
\item In vacuum the quadratic form $_g S_a^{~b}{}v^a v_b$, where $v^av_a = 1$, is {\it
positive-definite} giving the gravitational {\it superenergy density} $\epsilon_g$
for a fiducial observer {\bf O}.
\item In general, the canonical superenergy tensors are uniquely
determined only along the world line of the observer {\bf O}. But
in special cases, e.g., in Schwarzschild spacetime or in Friedman
universes, when there exists a physically and geometrically
distinguished four-velocity $v^i(x)$, one can introduce in an
unique way the unambiguous fields $_g S_i^{~k}(x;v^l)$ and $_m
S_i^{~k}(x;v^l)$.
\item We have proposed in our previous papers to use the tensor $_g S_i^{~k}(P;v^l)$
as a substitute of the non-existing gravitational energy-momentum
tensor.
\item It can easily seen that the superenegy densities
$\epsilon_g := _g S_i^{~k}v^iv_k, ~~\epsilon_m := _m S_i^{~k}v^i v_k$
for an observer {\bf O} who has the four-velocity $v^i$ correspond
exactly to the {\it energy of acceleration} ${1\over 2}m {\vec a}{\vec a}$
which is fundamental in Appel's approach to classical mechanics
\cite{Appel}.
\end{enumerate}

In past we have used the canonical superenergy tensors $_g S_i^{~k}$
and $_m S_i^{~k}$ to local (and also, in some cases, to global)
analysis of well-known solutions to the Einstein equations like
Schwarzschild and Kerr solutions; Friedman and Goedel universes,
and Kasner and Bianchi I, II universes.
The obtained results were interesting (See \cite {Gar1}).

We have also studied the transformational rules for the canonical
superenrgy tensors under conformal rescalling of the metric
$g_{ik}(x)$\cite{Gar1,Gar2}.

The idea of the superenrgy tensors can be extended on angular
momentum also \cite{Gar1}.The obtained angular superenergy tensors
do not depend on a radius vector and they depend only on {\it
spinorial part} of the suitable gravitational angular momentum
pseudotensor.\footnote{We have used in our investigation the
Bergmann-Thomson expression on angular momentum in general relativity.}

\section{Stability of the solutions to the Einstein equations and
canonical superenergy tensors}
Recently we have observed an exciting correlation between the total superenergy density,
 $\epsilon_s := \epsilon_m + \epsilon_g $, and stability of
 solutions to the Einstein equations. Namely, we have noticed that
 when a solution is stable, then $\epsilon_s\geq 0$, and when the
 solution is  unstable, then $\epsilon_s <0$.
 \begin{center}
 The examples
 \end{center}
 \begin{enumerate}
 \item Exterior Schwarzschild ------ stable ------- $\epsilon_s >0$:
 \item Einstein static universe ---- unstable ----- $\epsilon_s<0$;
 \item Kerr solution --------------- stable ------ $\epsilon_s>0$;
 \item de Sitter universe ---------- unstable ---- $\epsilon_s<0$;
 \item Anti-de Sitter universe ----- unstable ---- $\epsilon_s <0$;
 \item Friedman universes ---------- stable ------ $\epsilon_s>0$;
 \item Bianchi I universe ---------- stable ------ $\epsilon >0$;
 \item Kasner universe ------------- stable ------ $\epsilon_s >0$;
 \item Exterior Reissner-Nordstroem - stable ----- $\epsilon_s >0$;
 \item Minkowski spacetime --------- stable ----- $\epsilon_s =0$.
 \end{enumerate}

Instability of the de Sitter and anti-de Sitter universes was
proved recently \cite{Garfin,Emel}.

One can easily see that the above mentioned correlation follows
from the Proposition.

{\bf Proposition} If a solution to the Einstein equations is stable, then $\epsilon_s
>0$, and if the solution is unstable, then $\epsilon_s <0$.

{\bf Proof}. $\star$ Our proof lies on the constructive definition (5). It is easily seen from it
that the sign of the superenergy density $S_a^{~b}(P) v^av_b =\star ~~S_0^{~0}(P)$
is determined by the sign of the integral in nominator because {\it (-) denominator
is always positive}.
Let us apply the definition (5) to the
component $K_0^{~0}(y)$ of the canonical energy-momentum complex, matter and gravitation.
This component gives us the total energy density, gravitation and matter, for
analyzed solution to the Einstein equations.

Let the analyzed solution will stable at point {\bf P}. Then, $K_0^{~0}(y)$ has minimum
at this point.

One can see from (5) and (12)that
$S_0^{~0}(P) = _g S_0^{~0}(P) + _m S_0^{~0}(P) = \epsilon_s (P)>0$ in the case because
$K_{(0)}^{~~~(0)}(y) - K_0^{~ 0} >0$. Contrary, if {\bf P} is an instability point of the
analyzed solution, then $K_0^{~0}(y)$ has maximum at this point, and it follows that
$S_0^{~0}(P) = _g S_0^{~0}(P) + _m S_0^{~0}(P) = \epsilon_s(P)<0$ because
$K_{(0)}^{~~~(0)}(y) - K_o^{~0}(P) <0$ in this case.

If the analyzed solution is stable (or unstable) in a 4-dimensional domain $\Omega$,
then one has in this domain $S_0^{~0}(y) = \epsilon_s(y) >0$
(or, respectively, $S_0^{~0}(y) = \epsilon_s(y) <0$).

The direct calculation shows that for the stable Minkowskian
spacetime  one has limiting case with value $\epsilon_s (y) =0$.$\star$

We also conjecture  that the inverse is correct: if for a solution to
the Einstein equations $\epsilon_s>0$, then the solution is
stable, and if for a solution $\epsilon_s<0$, then this solution
is unstable.

The argumentation which supports the conjecture is the following: Minkowski
spacetime which has $\epsilon_s =0$ is stable solution to the
Einstein equations. On the strength of that one can
conclude that the same should be correct in all the
cases in which $\epsilon_s>0$. (In this case the {\it stabilizing}
positive differences $K_{(0)}^{~~~(0)}(y) - K_0^{~0}(P)$ in the formula
(5) overbalance {\it destabilizing} negative differences
$K_{(0)}^{~~~(0)}(y) - K_0^{~0}(P)$). On the other hand, if $\epsilon_s<0$, then the
{\it destabilizing} negative differences $K_{(0)}^{~~~(0)}(y) - K_0^{~0}(P)$
overbalance stabilizing positive differences $K_{(0)}^{~~~(0)}(y) - K_0^{~0}(P)$
and, in the case, the analyzed solution should be unstable.

\section{Conclusion}
On the {\it superenergy level} we have no problem with suitable tensor for
gravity, e.g., one can introduce gravitational {\it canonical superenergy
tensor}.
The canonical superenergy tensors, gravitation and matter, are useful to local analysis of
the solutions to the Einstein equations, especially to analyze
of their singularities\cite{Gar1}.

In this paper we have proposed an application of these tensors  to study stability of
the solution to the Einstein equations.

We think that this new application of the superenergy tensors can be very useful.

\acknowledgments
This paper was mainly supported by Institute of Mathematics, University of Szczecin
(Grant No 503-4000-230351).

\section{Appendix}
We give here the canonical superenergy densities $\epsilon_s$ for
de Sitter, anti-de Sitter, static Einstein and Reissner-
Nordst\"om spacetimes. For simplicity we will use in here the {\it
geometrized units} in which $G = c =1$.

\begin{enumerate}
\item De Sitter spacetime ----- $\epsilon_s = (-) {28\over 27}\alpha \Lambda^2
<0$;
\item Anti-de Sitter spacetime ---- $\epsilon_s = (-) {32\over 27}\alpha\Lambda^2
<0$;
\item Einstein static universe ---- $\epsilon_s = (-){4\alpha\over
3R^4} <0$, where ${1\over R^2} = 4\pi\bigl(\rho +p\bigr)~= ~\Lambda - 8\pi p >0$;
\item Exterior Reissner-Nordstr\"om spacetime ----
\begin{eqnarray}
\epsilon_s &=& {2\alpha\over 9r^8}\bigl[3\bigl(2Q^2-r_s r\bigr)^2 +
5\bigl(Q^2 - r_s r\bigr)^2 + 2\bigl(3Q^2-r_s r\bigr)^2\nonumber\\
&+& 2\bigl(3Q^2 - r_sr\bigr)\bigl(2Q^2 - r_sr\bigr)\bigr]\nonumber\\
&+& {2Q^2\over r^8}\bigl(r_sr-2Q^2\bigr) + {12 Q^2\Lambda_{RN}\over
r^6}.
\end{eqnarray}
The last expression is positive for $r\geq r_H = m + \sqrt{m^2 -
Q^2}$, i.e., outside and on horizon $H$ of the Reissner-Nordstr\"om black hole.
\end{enumerate}

Here $r_s := 2m, ~~\Lambda_{RN} := 1 - {2m\over r} + {Q^2\over
r^2}$, and $m^2>Q^2$.

The total superenegy densities for the other solutions to the
Einstein equations mentioned in this paper have been  already
given in past \cite{Gar1}.

\end{document}